\documentclass[a4paper,12pt]{article}
\pdfoutput=1 

\usepackage{a4wide}
\usepackage{cite}

\usepackage[T1]{fontenc} 
\usepackage[latin1]{inputenc}

\usepackage{empheq}
\usepackage{amssymb}
\usepackage{amsmath}
\usepackage{amsfonts}
\usepackage{xcolor}

\DeclareMathAlphabet{\mathpzc}{OT1}{pzc}{m}{it}

\numberwithin{equation}{section}

\usepackage[colorlinks]{hyperref}

\hypersetup{
 citecolor=blue,
 linkcolor=blue,
 urlcolor=blue}

\newcommand{\be}{\begin{equation}}
\newcommand{\ee}{\end{equation}}
\newcommand{\bes}{\begin{equation}\begin{split}}
\newcommand{\ees}{\end{equation}\end{split}}

\newcommand{\ca}{c_{\alpha}}
\newcommand{\sa}{s_{\alpha}}
\newcommand{\mix}{X_{H/A}}
\newcommand{\Dphill}{\Delta \phi_{l_1\bar{l}_2}}
\newcommand{\cth}{\cos\theta_1}

\begin{document}
\vspace{-5.0cm}
\begin{flushright}
CERN-TH-2019-198, TTP19-040, P3H-19-046
\end{flushright}

\vspace{2.0cm}

\begin{center}
{\large \bf 
  Mixed scalar-pseudoscalar Higgs boson production through next-to-next-to-leading order at the LHC}\\
\end{center}

\vspace{0.5cm}

\begin{center}
Matthieu Jaquier$^{1}$, Raoul R\"ontsch$^{2}$.\\
\vspace{.3cm}
{\it
  {}$^1$Institute for Theoretical Particle Physics, KIT, Karlsruhe, Germany\\
  {}$^2$Theoretical Physics Department, CERN, 1211 Geneva 23, Switzerland\\~\\  
}
{\it E-mail}: matthieu.jaquier@kit.edu, raoul.rontsch@cern.ch
\vspace{1.3cm}

\end{center}

{\bf \large Abstract:}
We study the production of a mixed scalar-pseudoscalar Higgs boson in gluon fusion at the LHC,
through next-to-next-to-leading order (NNLO) in QCD.
We obtain fully differential results, including the decay of the Higgs boson to two charged lepton pairs.
We discuss the impact of the interference between the scalar and pseudoscalar states.
We also show differential distributions for several kinematic variables whose shape is sensitive to the
parity of the Higgs boson, and assess the impact of the NNLO QCD corrections on these shapes.

\section{Introduction}
\label{sec:intro}
Comprehensive studies of the properties of the Higgs boson will be a focus of the
particle physics community in the forthcoming years.
The large dataset that will be recorded by the Large Hadron Collider (LHC) during Run 3 and in its
high-luminosity phase will allow precise probes of the quantum numbers and couplings of the Higgs boson.
One of the most interesting properties of the Higgs boson is its parity.
Initial measurements indicate that the 125 GeV Higgs boson is a scalar state $J^P = 0^+$,
while the pseudoscalar state $J^P = 0^-$ has been ruled out~\cite{Chatrchyan:2012jja,Khachatryan:2014kca,Aad:2015rwa,Aad:2015mxa}.
On the other hand, the possibility that the Higgs boson is an admixture of scalar and pseudoscalar states has not been excluded yet by measurements,
although constraints on the parameter space for such a mixed state exist \cite{Sirunyan:2017tqd,Aaboud:2017vzb}.

Such mixing between scalar and pseudoscalar states implies $CP$ violation in the Higgs sector.
This would be a signal for physics Beyond the Standard Model (BSM),
and might explain the origin of $CP$ violation within the Standard Model (SM).
Moreover it could shed light on the physics of the early universe, 
given the  large enough amount of $CP$ violation that is required for baryogenesis. 
However, we also stress that, irrespective of explicit BSM scenarios, determining the behavior of the Higgs boson under parity is important
as  a matter of principle, 
in order to build a complete picture of this new particle.

From a phenomenological point of view, the possibility of producing mixed scalar-pseudoscalar Higgs states through gluon fusion
has been considered in Refs.~\cite{Gao:2010qx,Artoisenet:2013puc,Maltoni:2013sma}.
These references studied a number of angular observables whose shapes are sensitive to the parity of the Higgs boson.
However, Ref.~\cite{Gao:2010qx} considers the leading-order (LO) production,
while Refs.~\cite{Artoisenet:2013puc,Maltoni:2013sma} include the next-to-leading order (NLO) QCD corrections and parton shower effects.
It is well known that Higgs production is subject to large perturbative QCD corrections and that results at
LO and NLO are not always reliable.
Indeed, the results for the scalar Higgs boson are known at next-to-next-to-next-to-leading order (N3LO) accuracy~\cite{Mistlberger:2018etf},
and these indicate that the perturbative expansion in $\alpha_s$ only begins to converge at next-to-next-to-leading order (NNLO). 
The NNLO corrections for both scalar and pseudoscalar Higgs boson production have been known for some time ~\cite{Anastasiou:2002yz,Harlander:2002vv,Anastasiou:2002wq},
and are implemented in the public codes {\tt SuSHi}~\cite{Harlander:2012pb,Harlander:2016hcx} and its extension {\tt SuSHiMi}~\cite{Liebler:2016ceh}.
These results, however, are limited in two ways: first, they do not include the possibility of an admixture between scalar and pseudoscalar Higgs states,
including possible interference effects.
Second, they only include the decays of the Higgs boson as overall branching ratios.
As a result, they cannot be used to
investigate the potential to determine the parity of the Higgs boson using
information from its decay products,
nor can they be used to compute fiducial cross sections defined by cuts on the decay products of the Higgs boson.

In this paper, we aim to bridge this gap by presenting the first NNLO QCD-accurate
fully differential predictions of mixed scalar-pseudoscalar Higgs boson production,
including the Higgs decay into two charged lepton pairs.
In particular, we  consider 
distributions in the angles $\Dphill$, $\Phi$, and $\cth$,
which describe the geometry of the decay 
of a spin zero particle into two charged lepton pairs and are known to be sensitive
to its parity  \cite{Gao:2010qx}, 
and we assess the impact of the NNLO QCD corrections on these observables.

The remainder of the paper is organized as follows.
In Section~\ref{sec:technical}, we briefly summarize the technical details involved in this calculation,
before presenting results in Section~\ref{sec:result}, and concluding in Section~\ref{sec:concl}.

\section{Technical Details}
\label{sec:technical}
In order to describe the production of an arbitrary mix of scalar $(0^+)$ and pseudoscalar $(0^-)$ Higgs states,
we make use of the Higgs Characterization model introduced in Ref.~\cite{Artoisenet:2013puc}.
The Lagrangian describing the interaction between a spin zero particle and two heavy fermions is 
\be
\mathcal{L} \supset - \sum_{f=t,b,\tau} \bar{\psi}_f \left( \ca \; \kappa_{Hff}  \; g_{Hff} + i \sa \; \kappa_{Aff} \; g_{Aff} \gamma_5 \right) \psi_f \mix,
\ee
where $\psi_f$ is a fermionic field of flavor $f$, $\mix$ is the Higgs field, $g_{Hff}$ and $g_{Aff}$ are the scalar and pseudoscalar Higgs couplings to the fermions respectively,
and we have used the notation $\ca =\cos(\alpha)$ and $\sa = \sin(\alpha)$.
The mixing between the scalar and pseudoscalar Higgs states is therefore controlled solely by the parameter $\alpha$, 
with the pure scalar and  pseudoscalar states  given by $\ca=1$ and $\ca=0$, respectively.
It was shown in Ref.~\cite{Artoisenet:2013puc} that this Lagrangian can be used to build an effective Lagrangian
for the interaction of a spin-0 mixed scalar-pseudoscalar state with vector bosons below a cutoff scale $\Lambda$.
In this paper, we will focus on the production of the Higgs boson through gluon fusion and its subsequent decay into two charged lepton pairs.
The relevant terms in the effective Lagrangian are
\bes
\mathcal{L}_{\rm eff} \supset & \Biggl\{ \ca \kappa_{\rm SM} \frac{1}{2} g_{HZZ} Z_{\mu} Z^{\mu} -\frac{1}{4}\bigl[ \ca \kappa_{H\gamma\gamma} g_{H\gamma\gamma} A_{\mu \nu}A^{\mu\nu} + \sa \kappa_{A\gamma\gamma} g_{A\gamma\gamma} A_{\mu \nu}\tilde{A}^{\mu\nu}  \bigr] \\
&-\frac{1}{2}\bigl[ \ca \kappa_{HZ\gamma} g_{HZ\gamma} Z_{\mu \nu}A^{\mu\nu} + \sa \kappa_{AZ\gamma} g_{AZ\gamma} Z_{\mu \nu}\tilde{A}^{\mu\nu}  \bigr] \\
&-\frac{1}{4}\bigl[ \ca \kappa_{Hgg} g_{Hgg} G_{\mu \nu}G^{\mu\nu} + \sa \kappa_{Agg} g_{Agg} G_{\mu \nu}\tilde{G}^{\mu\nu}  \bigr] \\
&-\frac{1}{4}\frac{1}{\Lambda}\bigl[ \ca \kappa_{HZZ} Z_{\mu \nu}Z^{\mu\nu} + \sa \kappa_{AZZ} Z_{\mu \nu}\tilde{Z}^{\mu\nu}  \bigr] \\
&-\frac{1}{\Lambda}\ca \bigl[ \kappa_{H\partial \gamma} Z_{\nu} \partial_{\mu} A^{\mu \nu} + \kappa_{H\partial Z} Z_{\nu} \partial_{\mu} Z^{\mu \nu} \bigr] \Biggr\} \mix,
\end{split}
\label{eq:Leff}
\ee
where the field strength tensors are defined as
\bes
V_{\mu \nu} &= \partial_{\mu} V_{\nu} - \partial_{\nu} V_{\mu}\quad (V=Z,A), \\
G^a_{\mu \nu} &= \partial_{\mu} V^a_{\nu} - \partial_{\nu} V^a_{\mu} +g_s f^{abc} G_{\mu}^b G_{\nu}^c,
\end{split}
\ee
and the dual tensor is
\be
\tilde{V}^{\mu\nu} = \frac{1}{2}\epsilon_{\mu \nu \rho \sigma} V^{\rho \sigma}.
\ee
The factors $\kappa_{HVV}$ and $\kappa_{AVV}$ ($V=g,Z,\gamma$) allow modifications of the (dimensionful) couplings $g_{HVV}$ and $g_{AVV}$.
We comment on the values of these couplings in the following section.

We now discuss a technical detail concerning the renormalization of the pseudoscalar amplitudes.
Neglecting the Yukawa interaction between the pseudoscalar Higgs boson and the light quarks, the interaction between the $0^-$ state and light QCD particles is mediated by a top quark loop only. In the heavy top quark limit, this loop can be integrated out, leading to the two effective operators \cite{Chetyrkin:1998mw},
\begin{equation}
 \mathcal{O}_1^B=G_a^{\mu\nu}\tilde{G}_{a,\mu\nu}=\epsilon_{\mu\nu\rho\sigma}G_a^{\mu\nu}G_a^{\rho\sigma}\ ,\qquad\mathcal{O}_2^B=\partial_{\mu}\left(\bar{\psi}\gamma^{\mu}\gamma_5\psi\right)\ ,
 \label{eq:Lpseudo}
\end{equation}
where we denote bare operators with $\mathcal{O}_i^B$.
The first operator describes the interaction of the pseudoscalar Higgs boson with gluons, and is present in Eq.~\eqref{eq:Leff}.
The second operator describes the loop-induced interaction with light quarks, and  first appears at NNLO in QCD.
These operators mix under renormalization as
\begin{equation}
 \mathcal{O}_i^R=\sum_{j=1}^2Z_{ij}\mathcal{O}_j^B\ ,
\end{equation}
where $\mathcal{O}_i^R$ is the renormalized operator. Note that $Z_{21}=0$, and $Z_{22}$ is determined in such a way as to preserve the absence of higher order corrections \cite{Adler:1969er} to the axial anomaly,
\begin{equation}
 \mathcal{O}_2^R=\frac{\alpha_s}{\pi}\frac{n_f}{8}\mathcal{O}_1^R\ ,
 \label{eq:axial}
\end{equation}
to all orders in perturbation theory.

In the limit of massless light quarks, contributions from the squared operator $\left(\mathcal{O}_2^B\right)^2$ vanish and the only contribution from the operator $\mathcal{O}_2^B$ comes from the interference $\mathcal{O}_1^B\mathcal{O}_2^B$, which is absorbed in the renormalized operator $\mathcal{O}_2^R$. This contribution has to be added to the two-loop amplitudes for pseudoscalar Higgs boson production~\cite{Harlander:2002vv,Ravindran:2004mb,Ahmed:2015qpa}, which we did by expressing it in terms of the leading order contribution of $\left(\mathcal{O}_1^R\right)^2$ using Eq.~\eqref{eq:axial}.

We will use Eq.~\eqref{eq:Leff} and Eq.~\eqref{eq:Lpseudo} to describe the process $pp \to \mix \to e^- e^+ \mu^- \mu^+$ to NNLO in QCD.
We will consider the Higgs boson to be produced onshell, so that the production and decay processes factorize.
We briefly discuss our implementation of these two processes below.

The production is governed by the $g_{Hgg}$ and $g_{Agg}$ terms in Eq.~\eqref{eq:Leff}.
To compute the NNLO corrections, we need the double-real, real-virtual and double-virtual amplitudes
for the production of a scalar and a pseudoscalar Higgs boson.
We take these from Refs.~\cite{Berger:2006sh,Ravindran:2004mb,Badger:2006us,Glover:2008ffa,Dixon:2004za,Dixon:2009uk,Ahmed:2015qpa}, and use the nested soft-collinear subtraction scheme~\cite{Caola:2017dug,max,maxtc,Caola:2019nzf,Caola:2019pfz}
to extract and remove the infrared singularities associated with these contributions.
We have checked our results for scalar and pseudoscalar Higgs boson production through NNLO
against the program {\tt SuSHi}~\cite{Harlander:2012pb,Harlander:2016hcx} and found full agreement.

As far as the decay is concerned, Eq.~\eqref{eq:Leff} implies decays $\mix \to Z/\gamma^* Z/\gamma^* \to e^- e^+ \mu^- \mu^+$.
We have checked our implementation of the matrix elements for the leading order production and decay,
$gg \to \mix \to Z/\gamma^* Z/\gamma^* \to e^- e^+ \mu^- \mu^+$,
against {\tt MadGraph}~\cite{Artoisenet:2013puc,Alwall:2014hca} and find excellent agreement.

\section{Results}
\label{sec:result}
We now present numerical results for the production of mixed scalar-pseudoscalar Higgs states.
We consider the production of a Higgs boson of mass $m_H=125$ GeV at the LHC operating at 13 TeV center-of-mass energy.
We use a factorization and renormalization scale $\mu = m_H/2$ throughout the paper, and estimate the scale uncertainty
by varying this scale by a factor of 2 in either direction~\cite{Anastasiou:2016cez,Mistlberger:2018etf}.
We use NNPDF3.0 NNLO parton distribution functions~\cite{Ball:2014uwa} for all results,
i.e. we compute LO and NLO cross sections with NNLO PDF's.
The Higgs vacuum expectation value is taken to be
$v^2 = (G_F\sqrt{2})^{-1}$, where the Fermi constant is $G_F = 1.16639 \times 10^{-5}~{\rm GeV}^{-2}$.
We choose the mass of the $Z$ boson to be $m_Z = 91.1876~{\rm GeV}$ and its width to be $\Gamma_Z = 2.4952~{\rm GeV}$.
We use the weak coupling  $g_W^2 = 4 \sqrt{2} m_W^2 G_F $ and the weak mixing angle  $\sin^2\theta_W = 1- m_W^2/m_Z^2$, 
with the mass of the $W$ boson chosen to be $m_W = 80.398~{\rm GeV}$. 
The top mass is required for the Wilson coefficient; we take it to be $m_t = 173.2~{\rm GeV}$.

We begin by considering fully inclusive Higgs boson production through gluon fusion, without including the decay of the Higgs boson.
Referring to Eq.~\eqref{eq:Leff}, it is clear that the relevant interaction terms are
$G_{\mu \nu}G^{\mu\nu}\mix$ and  $G_{\mu \nu}\tilde{G}^{\mu\nu}\mix$, which are controlled by five parameters:
the dimensionful couplings $g_{Hgg}$ and $g_{Agg}$, the dimensionless parameters $\kappa_{Hgg}$ and $\kappa_{Agg}$ which allow the modifications of the couplings,
and the scalar-pseudoscalar mixing parameter $\ca$. The dimensionful couplings have the values
\be
g_{Hgg} = -\frac{\alpha_s}{3\pi v};\;\;\;\;\;\;\;\;\;\;g_{Agg} = \frac{\alpha_s}{2\pi v}.
\ee
We set $\kappa_{Hgg} = \kappa_{Agg} = 1$, and present results for three representative values of the mixing parameter, 
$\ca=\{1,0,\sqrt{1/2}\}$, which correspond to a pure scalar, pure pseudoscalar,
and an equal scalar-pseudoscalar admixture, respectively.

\begin{table}
\centering
\begin{tabular}{|l|c|c|c|}
\hline
  & $\sigma^{\rm LO} $ [pb]  & $\sigma^{\rm NLO} $ [pb] & $\sigma^{\rm NNLO} $  [pb] \\
\hline
\hline
$\ca=1 $             & $15.13^{-14\%}_{+16\%}$     & $34.81^{-14\%}_{+20\%}$     & $43.85^{-9\%}_{+9\%}$       \\
\hline
$\ca=0$              & $34.04^{-14\%}_{+16\%}$     & $79.01^{-15\%}_{+20\%}$     & $99.46^{-9\%}_{+9\%}$  \\
\hline
$\ca=\sqrt{1/2}$     & $24.59^{-14\%}_{+16\%}$     & $56.91^{-15\%}_{+20\%}$     & $71.66^{-9\%}_{+9\%}$     \\
\hline
\end{tabular}
\caption{Total inclusive cross sections for Higgs boson production at LO, NLO and NNLO at the 13 TeV LHC, for three values of $\ca$.
  The cross section is shown for the central scale choice $\mu=m_H/2$.
  The subscripts (superscripts) indicate the scale variation obtained by varying by a factor of 1/2 (2).
  See text for further details.
\label{tab:inclxsecs}
}
\end{table}

The cross sections at LO, NLO and NNLO in QCD are shown in Table~\ref{tab:inclxsecs}.
As is well known for Higgs boson production, the NLO and NNLO corrections are large, with NLO and NNLO $k$-factors of approximately
2.3 and 1.25, respectively, while the scale uncertainties at LO and NLO underestimate the missing higher order corrections.
We also note that the cross sections in the scalar case are smaller than those in the pseudoscalar case by a factor of about $0.44$,
due to the coupling of gluons to a scalar Higgs boson being suppressed by a factor $g_{Hgg}^2/g_{Agg}^2 = 4/9$ relative to the pseudoscalar Higgs boson.
We note that at all three orders, the result for $\ca=\sqrt{1/2}$ is the arithmetic average of the results for $\ca=0$ and $\ca=1$.
This implies that there are no interference contributions $\sim \ca \sa$.
This is immediately obvious at LO and NLO as the relevant matrix elements do not include such interference terms.
However, both the double-real and the real-virtual matrix elements which enter the NNLO calculation contain such terms $\sim \ca \sa$, which only vanish upon integration
over the phase space.
We have confirmed this observation for the case of LO gluon fusion Higgs production in association with two jets
(which corresponds to the fully resolved double-real contributions to the NNLO corrections), both using our own code and using {\tt MadGraph}~\cite{Artoisenet:2013puc,Alwall:2014hca}.
  We therefore conclude that, if one considers the production of a Higgs boson and neglects its decay,
  the results up to NNLO for an arbitrary value of $\ca$ may be obtained by simply rescaling the scalar and pseudoscalar results
\be
\sigma(\ca) = \ca^2 \cdot \sigma(\ca=1) + \sa^2 \cdot \sigma(\ca=0).
\label{eq:reweight}
\ee

\begin{table}
\centering
\begin{tabular}{|l|c|c|c|}
\hline
                     & $\sigma^{\rm LO} $ [ab]  & $\sigma^{\rm NLO} $ [ab] & $\sigma^{\rm NNLO} $  [ab] \\
\hline
\hline
$\ca=1 $             & $10.6^{-14\%}_{+15\%}$     & $23.5^{-14\%}_{+19\%}$     & $29.1^{-8\%}_{+8\%}$       \\
\hline
$\ca=0$              & $0.0151^{-14\%}_{+15\%}$    & $0.0344^{-14\%}_{+19\%}$      & $0.0428^{-8\%}_{+8\%}$     \\
\hline
$\ca=\sqrt{1/2}$     & $8.61^{-14\%}_{+15\%}$      & $19.2^{-14\%}_{+19\%}$     & $23.7^{-8\%}_{+8\%}$  \\
\hline
$\ca=0.6$            & $9.95^{-14\%}_{+15\%}$     & $22.4^{-14\%}_{+19\%}$     & $27.7^{+8\%}_{-8\%}$  \\
\hline
\end{tabular}
\caption{Fiducial cross sections for $pp \to H \to ZZ^* \to e^- e^+ \mu^- \mu^+$ at LO, NLO and NNLO at the 13 TeV LHC, for four values of $\ca$.
  The cross section is shown for the central scale choice $\mu=m_H/2$.
  The subscripts (superscripts) indicate the scale variation obtained by varying by a factor of 1/2 (2).
  The kinematic cuts and parameter choices are described in the text.
\label{tab:fidxsecs}
}
\end{table}

We now turn to the case of the Higgs boson decaying into two charged lepton pairs $pp \to \mix \to Z/\gamma^*  Z/\gamma^*  \to e^-e^+ \mu^- \mu^+$, with the Higgs boson being onshell.
We impose minimal kinematic cuts on the final state leptons, inspired by a recent ATLAS analysis \cite{Aaboud:2017vzb}.
We require all leptons to have transverse momentum $p_{T,l} > 15$ GeV and pseudorapidity $|\eta_{l} | < 2.5$.
Moreover, we require the invariant mass of each lepton pair to be in a window around the $Z$ mass peak, $50~{\rm GeV} < m_{l^- l^+} < 106~{\rm GeV}$.
This last cut implies that the contribution of the offshell photons is negligible;
for simplicity, we set $\kappa_{H\gamma\gamma} = \kappa_{A\gamma\gamma} = \kappa_{HZ\gamma} = \kappa_{AZ\gamma} = 0$ in Eq.~\eqref{eq:Leff}.
Moreover, since we are interested in $CP$-violation in the Higgs sector, we will set $\kappa_{H\partial Z}=\kappa_{H\partial \gamma}=0$
as these derivative terms do not have a pseudoscalar counterpart.
Therefore, we will only consider the terms in Eq.~\eqref{eq:Leff} which are governed by $\kappa_{HZZ}$ and $\kappa_{AZZ}$,
together with the SM term with coupling $g_{HZZ}=2m_Z^2/v$, and the production terms with couplings $g_{Hgg}$ and $g_{Agg}$ which we have already discussed.
We then set $\kappa_{\rm SM} = \kappa_{HZZ} = 1$, $\kappa_{AZZ} = 1$, $\Lambda = 1$ TeV, and consider the three benchmark scenarios with  values of $\ca=\{0,1,\sqrt{1/2}\}$.
We also consider the case $\ca=0.6$ together with  $\kappa_{AZZ} = 20$  in order to illustrate the sensitivity of 
the shape information of certain angular observables to  the parity of the Higgs boson.
All other choices of parameters, scales, and the pdf set are the same as for the undecayed Higgs boson, described above.
\begin{figure}
 \centering
 \includegraphics[scale=1.0]{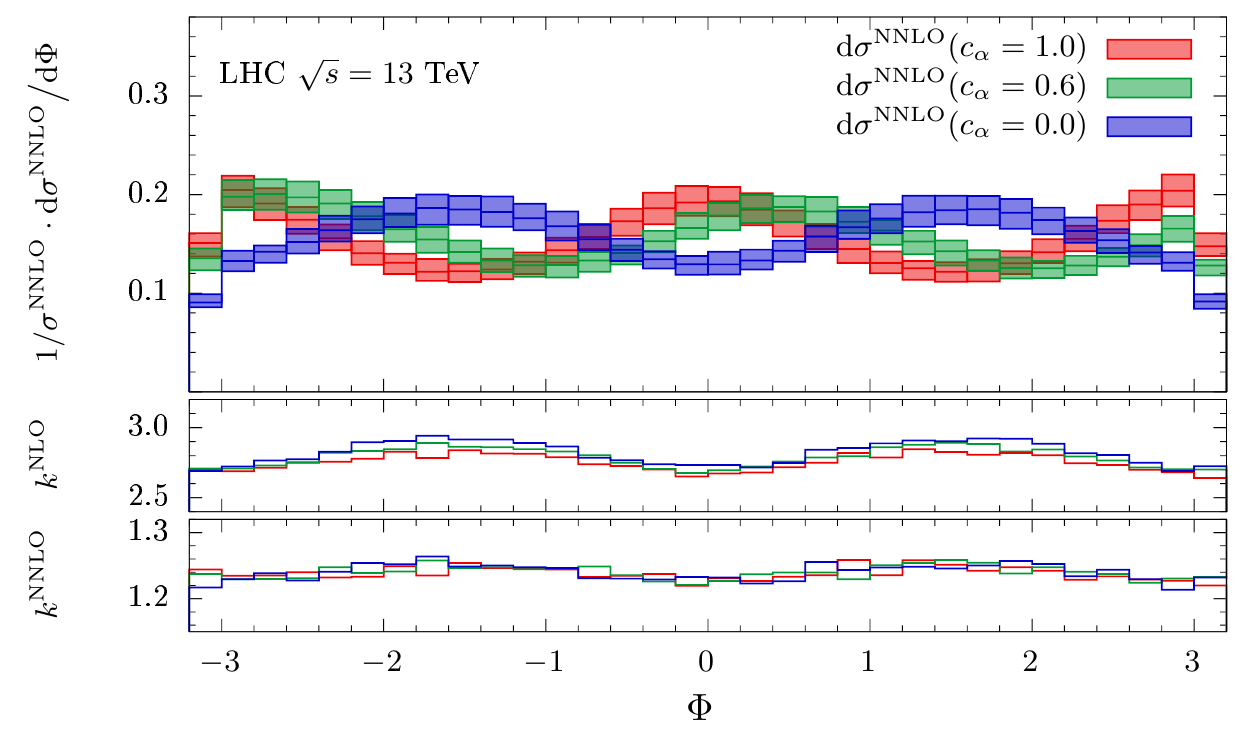}
 \caption{Normalized distribution at NNLO accuracy and  differential $k$-factors  $\mathrm{d}k^{\rm NLO}$ and $\mathrm{d}k^{\rm NNLO}$ for the angle $\Phi$. See text for further details.}\label{fig:phi}
\end{figure}

We show the fiducial cross sections for this setup  in Table~\ref{tab:fidxsecs}.
We first note that, in contrast to the results presented in Table~\ref{tab:inclxsecs},
the cross sections for the pure pseudoscalar case are smaller than those for the pure scalar case by three orders of magnitude.
This can be understood by looking at Eq.~\eqref{eq:Leff}.
The (scalar) SM interaction between the Higgs boson and the $Z$ boson pair has a coupling given by $g_{HZZ}=2m_Z^2/v$, as mentioned previously.
The pseudoscalar interaction $Z_{\mu \nu} \tilde{Z}^{\mu \nu} \mix$ leads to a factor $f(\{p\}) / \Lambda$ in the decay amplitude,
where $f(\{p\})$ is a kinematic factor with dimension of mass-squared. 
The value of $f(\{p\})$ is generally smaller than $m_Z^2$,
and moreover $\Lambda > v$, leading to the pseudoscalar decay to $Z$ bosons being suppressed relative to the SM scalar decay
by several orders of magnitude.

From Table~\ref{tab:fidxsecs} one can also see that the fiducial cross section for  $\ca=\sqrt{1/2}$
is no longer given by an arithmetic average of the fiducial cross sections for $\ca=1$ and $\ca=0$.
This is clear from the fact that the degree of scalar-pseudoscalar mixing is controlled by $\ca$
both in the production as well as in the decay.
This implies that terms $\sim \ca \sa$ do appear -- most notably,
from the combination of the pseudoscalar interaction in production and SM  interaction in the decay.
This means that, in general, a simple reweighting formula like
Eq.~\eqref{eq:reweight} cannot be used anymore due to the interplay between the
production and decay of the Higgs boson.

We note that the scale uncertainties and the impact of the NLO and NNLO corrections
are similar to those for the undecayed results (see Table~\ref{tab:inclxsecs}).
Moreover, both the scale uncertainties and the effects of the NLO and NNLO corrections are the same for all four values of $\ca$ in Table~\ref{tab:fidxsecs}.
This, together with the fact that for the SM Higgs, the N3LO corrections lie within the NNLO scale uncertainty bands~\cite{Anastasiou:2016cez,Mistlberger:2018etf},
lead us to conclude that NNLO is the first order at which the results for any value of $\ca$ are reliable.

It is clear that, for this choice of parameters, the cross sections provide enough information to discriminate
between the pure scalar and pure pseudoscalar scenarios.
If we compare the results for the $\ca=1$ and $\ca=\sqrt{1/2}$ cases,
we see that they are compatible within the scale uncertainties at LO and NLO.
The NNLO corrections, however, lead to reduced scale uncertainties,
and the results for these two values of $\ca$ are no longer compatible at this order.
This  emphasizes the need for higher order corrections in determining the properties of the Higgs boson.
On the other hand, the results for $\ca=1$ and $\ca=0.6$ (with $\kappa_{AZZ} = 20$) are compatible within the scale uncertainties at LO, NLO and NNLO,
meaning that one cannot differentiate between these two cases based on the rates alone,
and additional information from the shape of kinematic distributions is required.

We now show differential distributions for three observables $\Phi$,  $\cth$  and  $\Dphill$.
The first two  observables were proposed in Ref.~\cite{Gao:2010qx}, where they have been shown to be par\-ti\-cu\-lar\-ly sensitive to the spin and parity of the Higgs boson.
The observable $\Phi$ is the azimuthal angle between the planes constructed by the respective decay products of the two $Z$ bosons in the rest frame of the Higgs boson, while
$\cth$ is the polar angle of the decay products of the first $Z$ boson in its own rest frame.
Identifying the $Z$ boson that decays to electrons as $Z_1$ and the one that decays into muons as $Z_2$, the angle $\Phi$ is defined as~\cite{Gao:2010qx,Bolognesi:2012mm}
\be
\Phi = \frac{\vec{q}_1 \cdot \left (\hat{n}_1 \times \hat{n}_2 \right)} { | \vec{q}_1 \cdot \left (\hat{n}_1 \times \hat{n}_2 \right) |} \times \arccos\left(-\hat{n}_1 \cdot \hat{n}_2 \right),
\ee
where $\vec{q}_i$ is the three-momentum of $Z_i$, 
\be
\hat{n}_i = \frac{\vec{q}_{i1} \times \vec{q}_{i2}}{|\vec{q}_{i1} \times \vec{q}_{i2}|},
\ee
and $\vec{q}_{i1(2)}$ is the three-momentum of the lepton (antilepton) resulting from the decay of $Z_i$.
Here, all three-momenta are defined in the rest frame of the Higgs boson.
The observable $\cth$ is defined as~\cite{Gao:2010qx,Bolognesi:2012mm}
\be
\cth = - \frac{\vec{q}_2 \cdot \vec{q}_{11}}{|\vec{q}_2| |\vec{q}_{11}|},
\ee
where  $q_2$ and $q_{11}$ are now defined in the rest frame of a scattering axis whose direction is that of $Z_1$ in the Higgs rest frame.

We begin by showing the $\Phi$ distribution in Fig~\ref{fig:phi}.
The upper pane shows the NNLO results  for $\ca=1.0$, $\ca=0.6$ (with $\kappa_{AZZ}=20$)  and $\ca=0.0$, normalized to their respective NNLO cross sections.
The distribution obtained for the scale $\mu=m_H/2$ is depicted as the thicker central line,
while the band around it is the envelope from varying the scale by factors of two and $1/2$ around this central value.
As expected from  Ref.~\cite{Gao:2010qx}, there is  a notable shape difference between the three values of $\ca$.
The value $\Phi=0$ corresponds to a maximum for the scalar distribution and a minimum for the pseudoscalar, while the scalar has minima and the pseudoscalar has maxima 
at $\Phi=\pm \pi/2$.
The minima and maxima of the $\ca=0.6$ distribution are shifted relative to the pure scalar and pure pseudoscalar cases, giving the $\ca=0.6$ case a distinct shape.
This shift originates from the interference between the $0^+$ and $0^-$ production and decay contributions.
We recall from Table~\ref{tab:fidxsecs} that one cannot tell the $\ca=1$ and $\ca=0.6$ scenarios apart based on overall rates alone, 
as  the cross sections for these values of $\ca$ lie within each others' uncertainty  bands, even once NNLO corrections are taken into account.
The shape of the $\Phi$ distribution may provide a means to distinguish between these two scenarios.
Of course, the choice of parameters $\ca=0.6$ and $\kappa_{AZZ}=20$ is somewhat contrived, but it does illustrate the importance of
shape information in determining the parity of the Higgs boson in a large EFT parameter space.

\begin{figure}
 \centering
 \includegraphics[scale=1.0]{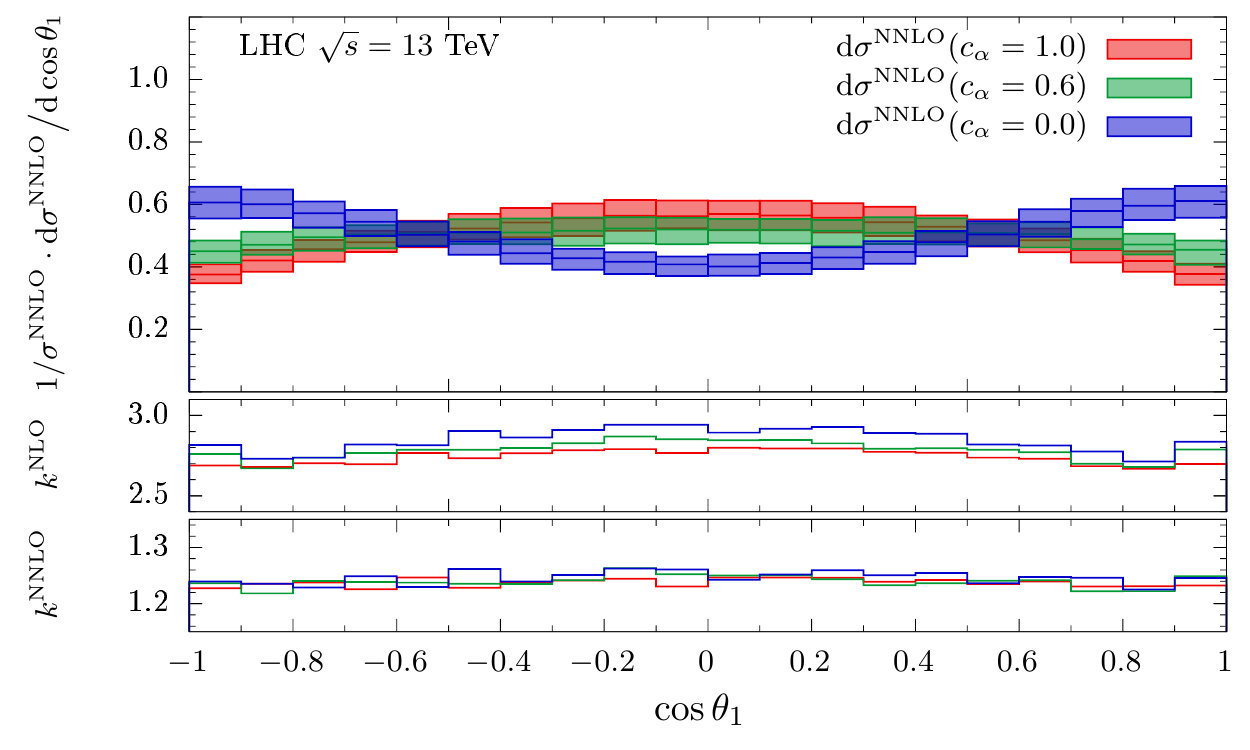}
 \caption{Normalized distribution at NNLO accuracy and  differential $k$-factors $\mathrm{d}k^{\rm NLO}$ and $\mathrm{d}k^{\rm NNLO}$ for the $\cos\theta_1$ observable.
   See text for further details.}\label{fig:costh}
\end{figure}

It follows from this discussion that it is important to have reliable predictions for the shapes of distributions,
meaning that the impact of the NNLO corrections needs to be known.
In the lower two panes of Fig.~\ref{fig:phi}, we show the differential NLO and NNLO $k$-factors, defined as
\begin{equation}
  {\rm d}k^{\rm NLO} = \frac{\mathrm{d}\sigma_{NLO}}{\mathrm{d}\sigma_{LO}}; \;\;\;\;\;\;
  {\rm d} k^{\rm NNLO} = \frac{\mathrm{d}\sigma_{NNLO}}{\mathrm{d}\sigma_{NLO}}.
\end{equation}
In order to simplify the plot, the $k$-factors are only shown for the central scale choice.
We observe that the value of $\Phi$ has a mild effect on the NLO $k$-factor, which peaks at $\Phi = \pm \pi/2$ and is smallest at $\Phi=0$,
and therefore tends to reduce the difference between the scalar and pseudoscalar distributions.
This behavior of the NLO $k$-factor appears to be an effect of the real radiation in conjunction with the cuts;
without the cuts, we observe the $k$-factor to be perfectly flat, as expected.
We conclude that the real emission moves the final state inside or outside of the fiducial volume defined by the kinematic cuts,
in a way which is similar for the scalar and the pseudoscalar cases.
The NNLO $k$-factor is quite flat, and amounts to a simple rescaling of the NLO results by a factor of approximately 1.25,
implying that the additional radiation present at NNLO does not dramatically change the acceptance rates for this fiducial volume.
We also note that the NLO $k$-factor is slightly larger for the pseudoscalar Higgs than for the scalar one,
while the NNLO $k$-factors are almost identical for all three values of $\ca$.

We now turn to the $\cth$ distribution, shown in Fig.~\ref{fig:costh}.
As for the $\Phi$ distribution, the shape of this distribution is significantly different for the pure scalar and pure pseudoscalar scenarios,
with the former having a maximum and the latter a minimum at $\cth=0$.
The shape difference between the pure scalar and the $\ca=0.6$ distributions is much milder,
implying that this observable is less sensitive to the parity of the Higgs than $\Phi$ is, given our setup.
The NLO corrections have a mild dependence on the value of $\cth$ and appear to have a slightly larger impact at low values of this angle.
The NNLO $k$-factor is flat, and the $k$-factors at both NLO and NNLO are the same for all values of $\ca$.

\begin{figure}
 \centering
 \includegraphics[scale=1.0]{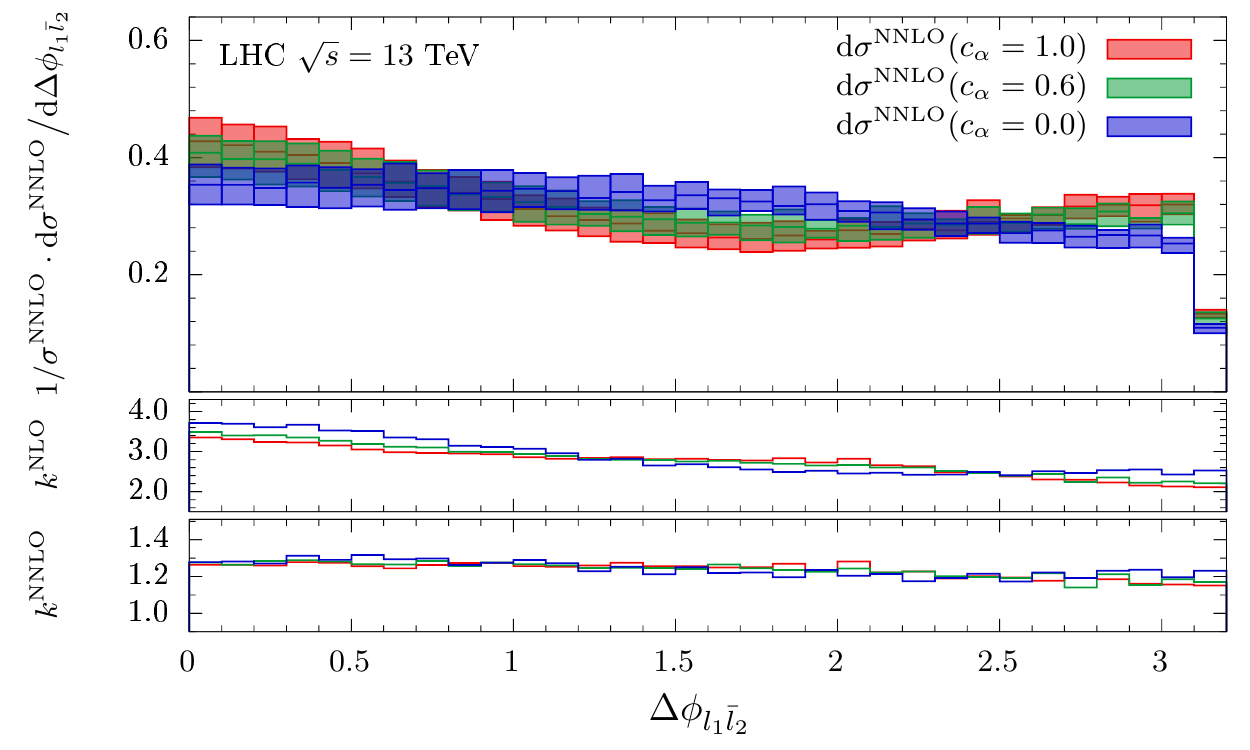}
 \caption{Normalized distribution at NNLO accuracy and differential $k$-factors $\mathrm{d}k^{\rm NLO}$ and $\mathrm{d}k^{\rm NNLO}$ for the $\Dphill$ observable.
 See text for further details.}\label{fig:deltaphi}
\end{figure}

Finally, we show the distribution in the opening angle $\Dphill$ in the lab frame between the $e^-$ and the $\mu^+$ leptons in Fig.~\ref{fig:deltaphi}.
Unlike the $\Phi$ and $\cth$ observables considered previously, this observable can be measured
even if the final state lepton configuration -- and thus the Higgs boson rest frame --  cannot be fully reconstructed.
As such, it is also an interesting proxy for the $W^+W^-$ decay channel of the Higgs boson.
We see a noticeable shape difference between the pure scalar and pure pseudoscalar cases,
but the difference between the $\ca=0.6$ and the pure scalar or pure pseudoscalar cases is much milder
and is covered by the scale uncertainty bands of the distributions.
Therefore, as expected, this observable has a lower sensitivity to the parity of the Higgs boson than either $\Phi$ or $\cth$.
Looking at the NLO $k$-factor, we see that the NLO corrections enhance the distribution at small angles.
This is due to the additional radiated parton, and the effect
is made more pronounced by the kinematic cuts that we impose.
On the other hand, the $k$-factor at NNLO is relatively flat,
implying that the presence of a second radiated parton has less of an impact, as we saw for the $\Phi$ and $\cth$ distributions.
Again, the value of $\ca$ does not seem to affect the differential $k$-factors at NLO or NNLO.

\section{Conclusions}
\label{sec:concl}
We have presented the first fully differential results for the production of a mixed scalar-pseudoscalar Higgs boson $\mix$ through gluon fusion to NNLO accuracy in QCD.
We made use of an effective Lagrangian to parametrize the coupling of the mixed state to gluons as well as its decay into $Z$ bosons.
In particular, the mixing between the scalar and pseudoscalar Higgs states is controlled by a single parameter $\ca$.
This allows us to make precise predictions for a generic observable at the LHC for an arbitrary admixture of scalar and pseudoscalar Higgs states.

For the production of a stable $\mix$ boson we observe that the cross section for an arbitrary mixing angle
can be obtained by an appropriate reweighting of the cross sections for scalar and pseudoscalar Higgs production.
This is true even at NNLO in QCD, where interference effects between the scalar and pseudoscalar amplitudes occur,
meaning that such interference vanishes upon integration over the full phase space.

Furthermore, we considered the subsequent decay $\mix\rightarrow ZZ^* \rightarrow e^-e^+\mu^-\mu^+$ for onshell intermediate $\mix$.
We observe the fiducial cross section for the pure pseudoscalar Higgs boson to be smaller than that for the pure scalar one by several orders of magnitude,
as a result of the pseudoscalar decay amplitudes being suppressed relative to the scalar ones.
This implies that the scalar and pseudoscalar bosons may be distinguished through the rates alone.
On the other hand, for certain values of the mixing parameter $\ca$ and of the coupling of the $Z$ bosons to the Higgs, the cross sections for the pure scalar and the mixed state are comparable.
In these situations, angular observables are known to provide additional discriminating power.
We considered three differential distributions, and found that for the setup considered here,
the angle $\Phi$ showed the most noticeable sensitivity to the parity of the Higgs.
We observed that, while the NLO corrections showed some dependence on the observable for these three distributions,
the $k$-factors for the NNLO corrections were relatively flat.
Furthermore, the corrections are largely independent of the value of $\ca$,
implying that the differential corrections to pure scalar production are a good approximation
for the differential corrections for any value of $\ca$.
However, one should be cautious in drawing this conclusion,
as the situation may change if different kinematic cuts are applied,
or if different values of the  parameters in the effective Lagrangian were chosen.

\vspace*{1cm}
{\bf Acknowledgments:}
We are grateful to Fabrizio Caola and Kirill Melnikov for helpful discussions.
We would also like to thank Stefan Liebner and Shruti Patel for explaining some details of the {\tt SuSHi} and {\tt SuSHiMi} codes, and
Ulascan Sarica and Markus Schulze for help with certain aspects of the {\tt JHUGen} code.
M.J. is supported by the Deutsche Forschungsgemeinschaft (DFG, German Research Foundation)
under grant 396021762 - TRR 257.

\end{document}